\documentclass[12pt,preprint]{aastex}
\usepackage{ulem}
\usepackage[usenames]{color}

\newcommand{\BP}{Ballesteros-Paredes}
\newcommand{\cs}{c_{\rm s}}

\newcommand{\kms}{{\rm ~km~s}^{-1}}

\newcommand{\Lj}{L_{\rm J}}

\newcommand{\Msun} {M_\sun}
\newcommand{\nmax}{n_{\rm max}}
\newcommand{\npk}{n_{\rm pk}}
\newcommand{\nsat}{n_{\rm sat}}
\newcommand{\nthr}{n_{\rm thr}}
\newcommand{\pcc}{{\rm ~cm}^{-3}}

\newcommand{\tad}{\tau_{\rm AD}}

\newcommand{\tff}{\tau_{\rm ff}}
\newcommand{\tpre}{\tau_{\rm pre}}
\newcommand{\tyso}{\tau_{\rm yso}}
\newcommand{\VS}{V\'azquez-Semadeni}

\slugcomment{Submitted to The Astrophysical Journal}

\shorttitle{CORE LIFETIMES IN NUMERICAL SIMULATIONS}
\shortauthors{GALV\'AN-MADRID ET AL.}

\begin{document}


\title{Statistics of Core Lifetimes in Numerical Simulations of
Turbulent, Magnetically Supercritical Molecular Clouds}


\author{Roberto Galv\'an-Madrid\altaffilmark{1}, Enrique
\VS\altaffilmark{1} Jongsoo Kim\altaffilmark{2}, and Javier
\BP\altaffilmark{1}}

\altaffiltext{1}{Centro de Radioastronom\'ia y Astrof\'isica (CRyA),
Universidad Nacional Aut\'onoma de M\'exico,
Apdo. Postal 72-3 (Xangari), Morelia, Michoac\'an 58089, M\'exico}
\email{r.galvan, e.vazquez, j.ballesteros@astrosmo.unam.mx}

\altaffiltext{2}{Korea Astronomy and Space Science Institute, 61-1,
Hwaam-dong, Yuseong-gu, Daejon 305-764, Korea}
\email{jskim@kasi.re.kr}


\begin{abstract}
We present measurements of the mean dense core lifetimes in numerical
simulations of magnetically supercritical, turbulent, isothermal
molecular clouds (MCs), in order to compare with observational
determinations. The mean ``prestellar'' lifetimes are given as a
function of the mean
density within the cores, which in turn is determined by the density
threshold $\nthr$ used to define them. The mean lifetimes are consistent
with observationally reported values, ranging from a few to several
free-fall times. We also present estimates of the
fraction of cores in the ``prestellar'',
 ``stellar'', and
``failed'' stages as a function of $\nthr$.  Failed cores are defined as
those that do not manage to collapse, but rather re-disperse back into
the environment. Due to resolution limitations, the
number ratios are
measured indirectly in the simulations, as either lifetime ratios (for
the prestellar cores), or as time-weighted mass ratios (for the failed
cores). Our approach contains
one free parameter, the lifetime of a protostellar object $\tyso$ (Class
0 + Class I stages), which is outside the realm of the
simulations. Assuming a value $\tau_\mathrm{yso} = 0.46$ Myr, we obtain
number ratios of starless to stellar cores ranging from 4--5 at $\nthr =
1.5 \times 10^4 \pcc$ to $\sim 1$ at $\nthr = 1.2 \times 10^5 \pcc$,
again in good agreement with observational determinations. We also find
that the failed cores are generally difficult to
detect, although the
mass in these cores is comparable to that in stellar cores at
$\nthr = 1.5 \times 10^4 \pcc$. At $\nthr = 1.2
\times 10^5 \pcc$ the mass in failed cores is negligible, in agreement
with recent observational suggestions
that at the latter densities the cores are in general gravitationally
dominated. We conclude by noting that the timescale for core contraction
and collapse is virtually the same in the subcritical, ambipolar
diffusion-mediated model of star formation, in the model of star
formation in turbulent supercritical clouds, and in a model
intermediate between the previous two, suggesting a convergence of the
models at least at the level of the core lifetimes, for currently
accepted values of the clouds' magnetic criticality.

\end{abstract}




\section{Introduction} \label{sec:intro}

The evolution of the dense cores within molecular clouds (MCs) on their route
to forming stars, and the duration of the various stages of contraction
and collapse in particular, are key ingredients in our understanding of
the star formation process. The process of contraction and final
collapse of a core is generally divided in the so-called ``prestellar''
phase, which includes the span between the time when the core is
first detected and when a protostellar object forms in its deepest
regions, and the ``stellar'' phase, which includes the Class 0 and Class
I protostellar stages (from the appearance of the protostellar object to
the clearance of the surrounding material).

Observationally, several works have estimated the duration of the
prestellar stage through either statistical or chemical methods
\citep[see the discussion in][ and references therein]{BKMV07}. The
former rely on measuring the number ratio of prestellar to stellar
cores, and assuming that this ratio equals the ratio of durations of
these stages \citep[e.g.,][]{Beich86, LM99, Oni02, Hat06, Jor07}. A
summary of prestellar durations, or ``lifetimes'', from
several studies, has been recently presented by \citet{WT_etal07}.

Theoretically, there are two main competing models that describe the
evolution of the cloud cores.  The so-called ``standard model'' of
quasi-static contraction mediated by ambipolar diffusion \citep[AD;
e.g.,][]{Mou76, Mou91, SAL87} was the dominant paradigm until
recently. It postulated that the cores in low-mass star-forming regions
evolved quasi-statically in magnetically subcritical clouds, contracting
gravitationally at the rate allowed by AD, which causes a redistribution
of the magnetic flux, until the central parts of the cores become
magnetically supercritical and proceed to dynamical collapse. The
process was originally thought to be slow, because the clouds were
thought to be strongly magnetically subcritical, in which case the AD
timescale $\tad$ is roughly one order of magnitude larger than the
free-fall time.  However, it was later recognized that near-magnetically
critical initial conditions \citep{FM93,CB01} or turbulent conditions
\citep{FA02, Heitsch_etal04} cause the AD timescale to become
significantly shorter, only a few times longer than the free-fall time
\citep[][ hereafter Paper I]{CB01, Paper I}. This
realization was motivated in part by observational studies suggesting
core lifetimes not much longer than their free-fall times
\citep[e.g.,][]{LM99, JMA99}. A recent numerical study 
of core lifetimes dominated by AD
but with turbulent initial conditions has been presented recently by
\citet{NL05}, finding core lifetimes in the range 1.5--10 times the
free-fall time.

On the other hand, the so called ``turbulent'' model of star formation
\cite[e.g.,][]{BVS99, VS_etal00, MK04, BKMV07} takes into account the
fact that MCs are turbulent, and attributes a dual character to the
turbulence: on the one hand it provides support against generalized
cloud collapse, but on the other it produces strong local density
enhancements, which constitute the clumps and cores within the clouds.
Thus, their formation timescales are of the order of the turbulent
crossing time across the size scales of the fluid parcels that collide
to produce the clumps and cores \citep{BHV99, Elm00, PN02}. However,
Paper I showed that the lifetimes of randomly selected cores at
densities comparable to those of observed ammonia cores are nevertheless
of the order of a few times the free-fall time of the cores, because
their evolution while they are detectable includes part of the assembly
process and then the collapsing stage. 
Note that this is contrary to
frequent perceptions that the turbulent model implies core lifetimes of
approximately \emph{one} free-fall time \citep[e.g.,][]{WT_etal07}.

Within the context of the turbulent model, it is natural to expect the
existence of an additional class of MC clumps that end up redispersing
rather than collapsing; that is, turbulent density fluctuations that do
not manage to become locally gravitationally unstable (Elmegreen 1993;
Taylor et al.\ 1996; Paper I). We refer to these as ``failed''
clumps. Also, in this model, it is not crucial whether the clouds are
magnetically subcritical or supercritical, since much of the ``filtering'' of
mass that reaches the collapsed objects, and which regulates the star
formation efficiency, is provided by the dual role of the turbulence.
A stronger magnetic field simply appears to provide an extra reduction
factor for the star formation efficiency, whose precise magnitude
depends on whether the turbulence in the clouds is driven or decaying
\citep[see the discussion in][ and references therein]{VS07}.

In this paper we present a survey of the core lifetimes in a set of numerical
simulations of a turbulent, continuously driven, magnetically
supercritical, isothermal MC, with the aim of obtaining statistically
reliable data about them within this model, that can be 
compared to
observational determinations. We also discuss the number ratio of
``starless'' to ``stellar'' (in the simulations, collapsed) cores. The
plan of the paper is as follows: In \S \ref{sec:simulation} we describe
the numerical simulations we analyzed; in \S
\ref{sec:procedure} we describe the method we used to measure the
lifetimes and the number ratios of the various kinds of objects; in  \S
\ref{sec:discussion} we compare them with existing observational and
theoretical work, and discuss their implications. In this section we also
discuss the limitations and accomplishments of the present work.

\section{The Numerical Simulations} \label{sec:simulation}

We have chosen the moderately supercritical simulation (named
M10J4$\beta$.1) reported in Paper I for analysis of the core
evolution. We refer to this run as ``R1'' in the rest of the paper,
and its selection was based on the fact that observations suggest that
MCs are nearly critical or moderately supercritical \citep{Crut99}.
The computational domain contained 64 Jeans masses, or 1860 $M_\odot$
\citep[][]{VKB05}. The mean number density is $n_0=500 \pcc$, the
simulation box size is 4 pc, and the (uniform) temperature is $T=11.4$
K, so that the sound speed is $\cs = 0.2 \kms$. The mean magnetic
field was $B_0 = 14.5~\mu$G, giving a mass-to-magnetic-flux ratio $\mu
=2.8$ times the critical value. These physical properties and
conditions of the simulation make it directly comparable to the most
massive ``supercores'' in the Perseus cloud, as defined by
\citet{Kirk06}. Indeed,
supercore No.\ 2 (IC 348) is reported by
those authors to have a mass $\sim 1940~\Msun$, and, at a distance of
250 pc, its size is $\sim 5$ pc, very close to the corresponding
parameters of our simulation. Supercore No.\ 7 (NGC 1333) is reported to be
somewhat smaller and less massive, at $\sim 970~\Msun$.

The turbulence in this simulation was continuously driven in order to
maintain a turbulent Mach number $\sim 10$.  For the turbulence random
driver, we follow the method in \citet{Stone98}, i.e., we applied
a random Fourier driver operating at the largest scales with the
functional form $P(k) \propto k^6 \exp{(-8k/k_{\rm peak})} $,  
where 
$k=2\pi/l$ is the wavenumber corresponding to the
scale $l$, $k_{\rm peak} = 2(2\pi/L)$ is the wavenumber where 
the input power spectrum
peaks, and $L$ is the one-dimensional size of the computational box.
As discussed in Paper I, this is motivated by the fact that turbulence
is ubiquitous at all scales in molecular clouds \citep[e.g.,][]
{Larson81, Blitz93, Heyer_Brunt04}, including starless ones such as
the so-called Maddalena's cloud \citep{Maddalena_Thaddeus85}, which
suggests a universal origin and maintenance mechanism for the clouds'
turbulence, and that turbulence is driven at the
largest scales in the clouds \citep{Heyer_Brunt04, Heyer_Brunt07}.

Two more simulations (called R2 and R3) were performed 
with the
same physical parameters but different random 
seeds for the Fourier driver. R2 
was performed at a resolution of 256 cells per dimension, and R3 
was performed at 512 cells per dimension. We used a total
variation diminishing (TVD) scheme \citep{Kim_etal99}. Runs R1 and R2
were integrated over 10~Myr, before self-gravity was turned on (5
turbulent crossing times). This is a standard procedure aimed at allowing the
system to attain a
well-developed, stationary turbulent 
state by the time gravity is turned on, thus preventing it from
``capturing'' features produced directly by the artificial random
driver. Note that, however, one turbulent crossing time is 
generally enough to 
achieve such a stationary state \citep{BP_etal06}. 
Thus, run R3 (the more expensive of
these simulations in terms of computational time) 
was run only for one dynamical time before self-gravity was turned on. 

Once self-gravity was turned on, runs R1, R2 and R3 were 
integrated over 13, 9.1, and 4.7 Myr respectively. 
In reality, however, as suggested by the estimated ages
of the stellar population in the Solar Neighborhood \citep{BHV99,
HBB01, BH07}, molecular clouds should 
live not much longer than 5 Myr,
since they are destroyed by stellar winds, bipolar outflows, SNe explotions, 
etc. For this reason, we only used the first 8 Myr of the evolution of runs 
R1 and R2 with self-gravity.

Note that run R1 was used to perform the analysis over all types of
cores (collapsed and failed).  However, since only two collapse events
were found in R1, we used runs R2 and R3 to verify that the timescales
for formation and collapse obtained in R1 were consistent, and
independent of the resolution.  Run R2 produced two collapse
events, while run R3 produced three.

\section{Measurement procedure} \label{sec:procedure}

\subsection{General considerations} \label{sec:gral_cons}

Just as with observations, estimating the core lifetimes in the
simulations is not a straightforward task. The numerical simulations
have both advantages and disadvantages compared to the
observations. First, in the simulations, 
the density, velocity
and magnetic fields defined over the three spatial and one temporal 
dimensions are given. 
However, the cores, defined as we describe below, are elusive entities 
that in general do
not preserve their identity over time; i.e., they do not involve the
same mass at different times. After all, defining a core amounts to
defining a certain region of space in what is really a fluid
continuum. We choose to define a core as the set of connected grid cells
surrounding a local density peak and having densities above a given
density threshold $\nthr$. This definition mimics the emission observed
when using tracers sensitive to different density regimes. We have
chosen not to use an algorithm such as CLUMPFIND 
\citep{Williams_etal94} 
because it can sometimes
introduce artificial divisions across mostly continuous objects. Our
chosen definition allows large-scale clumps to be
defined at low $\nthr$ even if the region contains more than one
local density maximum.

\begin{figure}
\epsscale{1.0}
\plotone{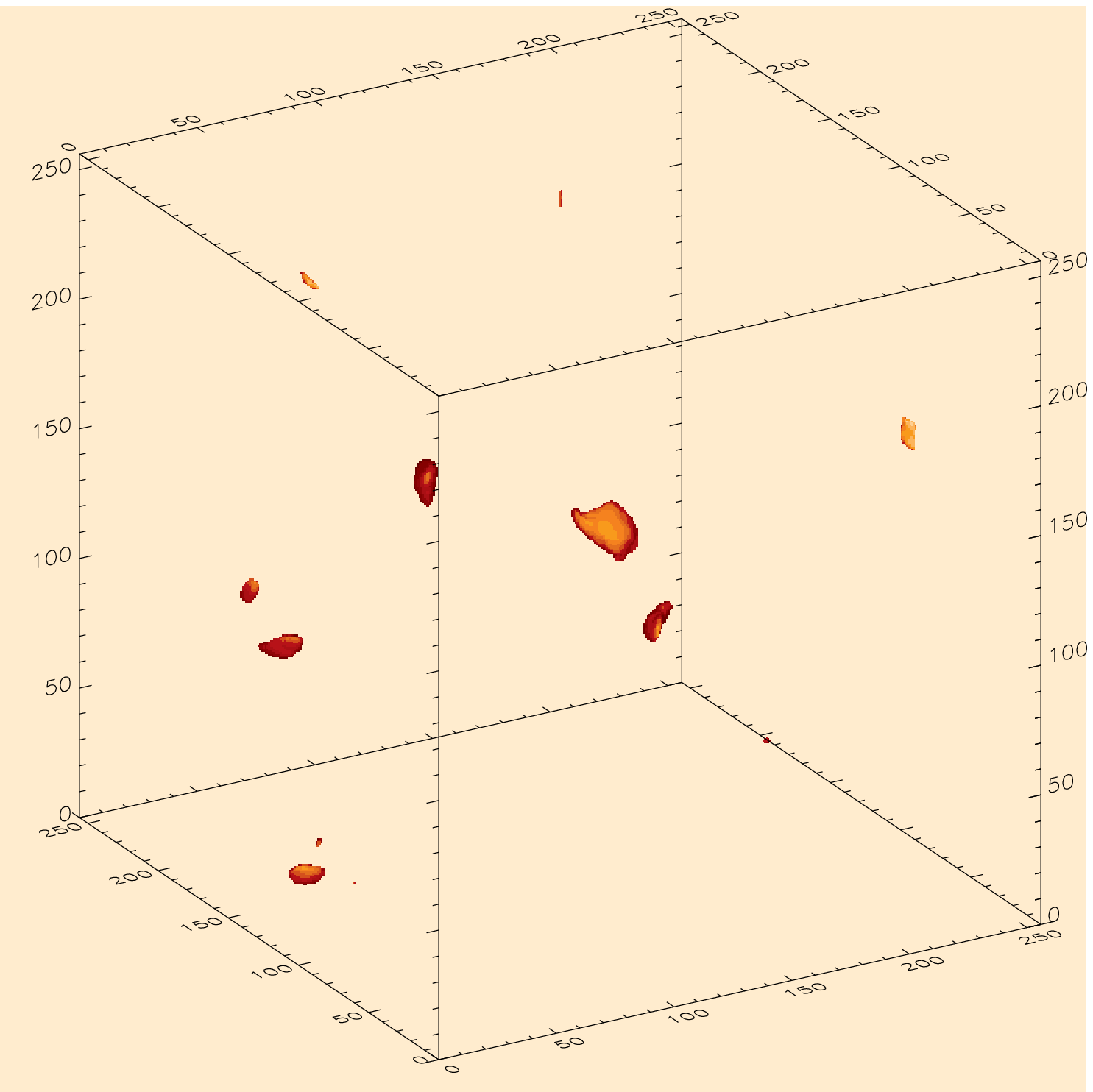}
\caption{Cores defined at $\nthr = 30~n_0 = 1.5 \times 10^4 \pcc$, and at
$t=1.96$ Myr for R1. This time marks the end of collapse for the first
collapsing core (the biggest core seen near the center of the simulation
box) in this run. At this threshold, failed cores become an important feature
and
many of them coexist at every temporal step. The two ``spots'' at
opposite sides of the $y$-axis boundaries, and at $(x,z) = (203,197)$
correspond to just one failed core crossing the periodic boundaries. An
animation of the simulation for $0.32 \le t/\mathrm{Myr} \le 8$ is
presented in the electronic version of the Journal, with the frames
separated by $\Delta t = 0.04$ Myr. The first 9 frames are omitted,
since there are no structures above $\nthr$ at those times. The figure
shown in the printed version of the Journal corresponds to frame \#40 in
the animation.}
\label{fig:core_evol}
\end{figure}

In general, the cores are more
poorly defined as lower values of $\nthr$ are considered, while instead this
problem disappears as one considers higher $\nthr$.
Once defined, the cores must be followed over
time. The animation of R1 associated to Fig.\ \ref{fig:core_evol}
in the electronic
edition shows the evolution of the cores when $\nthr = 30~n_0 =
1.5 \times 10^4 \pcc$, illustrating how the failed cores appear and
disappear, and sometimes merge or split. It also illustrates a merger of
collapsing cores (cf.\ \S \ref{sec:collapsing}).
The measured core properties (e.g., volume $V$, mass $m$, and mean
density $\bar{n}$) depend on $\nthr$, and furthermore, at a fixed value
of $\nthr$, these properties vary in time for every core.
Note in particular that the cores are not generally composed of the same
fluid parcels throughout their history.
Peak densities and their associated coordinates within a core
do not depend on $\nthr$, which only defines the time at which these two
properties begin to be measured.

The cores in the simulations can either collapse or rebound (no
stable hydrostatic solutions are possible in the isothermal, magnetically
supercritical regime; see Paper I). Thus, there exist two classes of
starless cores: those that are on route to collapse, to which we refer
as ``prestellar'', and those that will eventually redisperse back to
the surrounding medium, to which we refer as ``failed'' cores. In
addition, we refer to cores that have already collapsed as ``stellar''.
With these considerations in mind, we see that the number ratio of
starless to stellar cores is given by
\begin{equation}
\frac{N_{\star\mathrm{less}}}{N_\star}=\frac{N_\mathrm{pre}} {N_\star}
+\frac{N_\mathrm{f}}{N_\star} ,
\label{eq:starless2stellar}
\end{equation}
where $N_{\star\mathrm{less}}$, $N_{\rm pre}$, $N_\star$ and $N_{\rm f}$
are respectively the numbers of starless, prestellar, stellar and failed
cores.

Our goal in the present paper is to determine the lifetimes of the cores
and their number ratios as a function of the threshold used to define
them or, almost equivalently, as a function of their mean densities. The
lifetimes that are reported in observational studies correspond
to the duration of the prestellar stage, and are 
normally estimated by 
measuring the number of cores in the prestellar and stellar stages, and
assuming that the ratio of their numbers equals the ratio of
their durations. However, it was shown in Paper I that the ratio of the
numbers is an upper limit to the ratio of durations if there exists a
population of failed cores that is not accounted for. 
Furthermore, as discussed in \citet{BH07}, potential 
source count incompleteness in several observational surveys also
contributes to make this an upper limit.

In principle, one 
could simply measure the numbers of the various types
of cores (failed, prestellar and stellar) in the simulations. However, a
number of limitations prevent us from doing so. We have found that the
collapsed, or stellar, cores in our simulations are clearly
under-resolved. Their masses are typically between 50 and 100 $\Msun$,
while the smallest and densest cores in, e.g., Perseus, 
have masses $\sim 1 \Msun$,
although they are clustered in groups of up to several tens of them
\citep[see, for example, Figs.\ 4 and 5 of][]{Kirk06}. 
Specifically, we have extracted the masses of the clustered submillimeter
(submm) cores in supercore No.\ 2 and supercore No.\ 7 from tables 1, 2
and 3 of \citet{Kirk06}. We find a total mass in these clustered submm
cores of  
$\sim 7~\Msun$ for supercore No.\ 2  and $\sim 32~\Msun$ for supercore
No.\ 7. In comparison, the masses at the time of peak density saturation
of each of the collapsed objects in R1 at $\nthr
= 1.2 \times 10^5 \pcc$ (emulating the density of the submm cores) are
$61~\Msun$ and $62~\Msun$ respectively. 
Thus, each of our
cores with collapsed objects corresponds most directly to a cluster of
protostellar cores.

Another important issue is that the number of objects fluctuates
statistically in time. This
suggests that we should consider appropriately weighted time averages in
order to obtain the most representative numbers. As mentioned, our analysis 
in R1 was 
performed over the first 8 Myr of the simulation after gravity was turned
on, except for the failed cores at the lowest $\nthr$, for which a
sufficiently large statistical sample was obtained considering only the
first 4 Myr (due to the large number of these objects). Analysis in R2 and
R3 consisted only of collapse time measurements taken over 
8 Myr and 5 Myr (the duration of R3) after gravity is turned on respectively.

The time interval
between outputs in the simulations was 0.04 Myr, and we used four different 
values of the
density threshold: $\nthr = 1.5 \times 10^4, ~3 \times 10^4, ~6
\times 10^4,$ and $1.2 \times 10^5 \pcc$. Higher threshold densities were not
used because cores with $n > 256~n_0 = 1.28 \times 10^5 \pcc$ are not
well resolved in all the simulations according to the Jeans condition
\citep{True_etal97}, as
explained in Paper I.
Thresholds lower than $20~n_0 = 10^4 \pcc$ were not investigated either,
because at such low densities the cores become impossible to follow in
time due to their poorly defined identities (see, for example, the
animations accompanying Paper I). In the next subsections we describe
how we estimate the numbers and lifetimes of the various kinds of cores.

\subsection{Collapsing cores} \label{sec:collapsing}

\subsubsection{Time scales} \label{sec:coll_timescales}

\begin{figure}
\epsscale{0.6}
\plotone{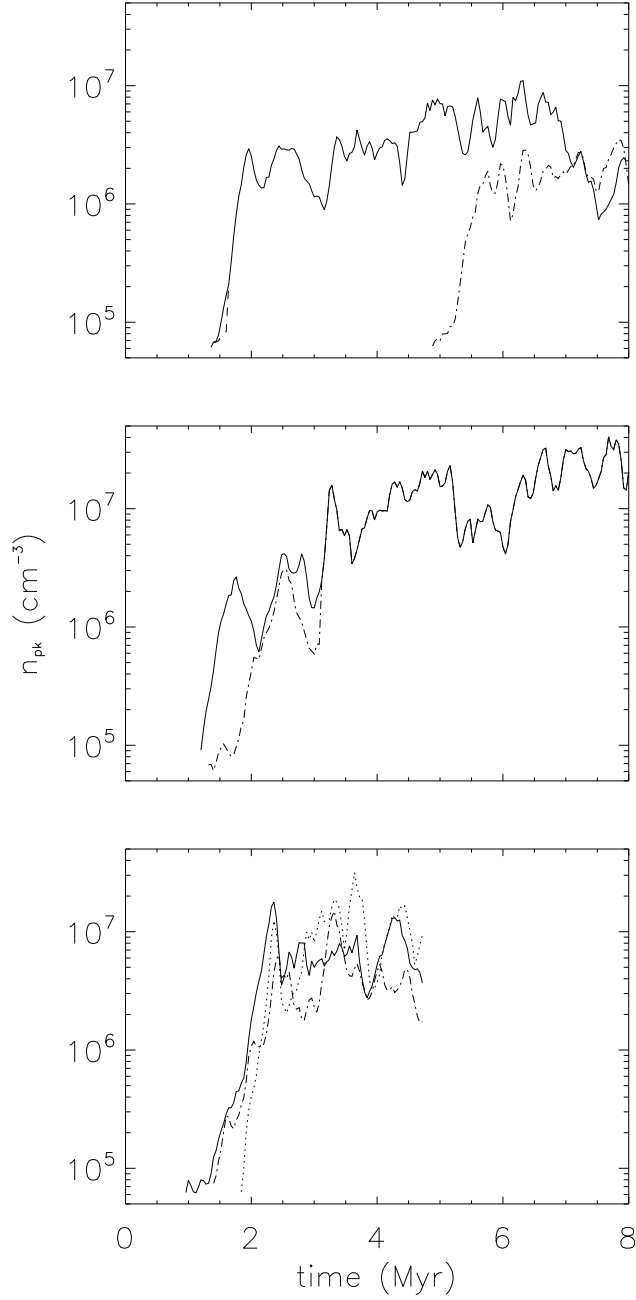}
\caption{Peak density as function of time for collapsing cores in R1 
({\it upper panel}), R2 ({\it middle panel}) and R3 ({\it lower panel})
after
they have exceeded $n_\mathrm{pk}=6\times10^4\pcc$.  A merging event
occurs among the first two collapsing cores of R1 ({\it solid} and {\it
dashed} lines respectively), but their peak densities are seen to rise
steeply before their merger (the time when the lines converge). 
A merger also occurs in R2, but well after the individual collapse 
of the cores.}
\label{fig:core_peak_dens_evol}
\end{figure}

When a core collapses in the simulations, in practice its peak
density $\npk$ does not increase to infinity, but rather reaches a
saturated value $\nsat \sim$ 2--3 $\times 10^6
\pcc$ at $256^3$ resolution or $\nsat \sim$ 1 $\times 10^7 \pcc$ in the $512^3$
simulation, beyond which the collapse cannot
be followed further by the code. 
In this state, the mass of the collapsed object
is spread out over a few pixels in each direction.
This process is
illustrated in Fig.\ \ref{fig:core_peak_dens_evol}, which shows the
evolution of the peak densities of the objects that collapse
in the three runs.
 Upper, middle and lower panels represent the peak density as a
 function of time for runs R1, R2, and R3, respectively. 
The saturated density of a collapsed
core in general oscillates in time. This is because the code contains no
prescription for ``capturing'' the mass in the collapsed objects (the
analogue of a ``sink particle'' in SPH codes, for example), and
therefore this mass still interacts with its surroundings. Also, $\nsat$
increases slowly on average, because 
the collapsed object continues to slowly accrete mass.

We define the \emph{prestellar} lifetime of a core, $\tpre$, as
the interval between the
 time when it is first detected at a given threshold $\nthr$, and the
 time when its peak density reaches the saturated value. This
 implicitly assumes that any further subfragmentation of the core
 occurs essentially simultaneously, so that all protostars form at roughly 
the same time, at least within the precision of our temporal resolution. This
definition of the prestellar lifetime neglects the
duration of the final stages of collapse (from
the time when the density saturates in the simulation to the time at
which actual protostellar densities would be reached). This
approximation should not introduce a large error, since the free-fall
time at the saturation densities is
\begin{equation}
\tff \equiv (3 \pi/32 G \rho)^{1/2} \sim 2 \times 10^{4} ~\hbox{yr},
\label{eq:free-fall_time}
\end{equation}
which is significantly
shorter than the temporal resolution of
$4 \times 10^{4}$ yr between snapshots in our simulation.
In any case, this slight underestimate of
the true prestellar lifetime acts in the opposite direction of any delay of
the collapse introduced by numerical diffusion, and thus the two effects
should tend to cancel each other out (see the 
discussion in \S \ref{sec:caveats}).

The measured prestellar lifetime obviously depends on $\nthr$, since this
quantity defines the starting time of a core's observed
``life''. Empirically, we have found that the mean density in the cores
scales close to linearly with $\nthr$ \citep[e.g.,][]{VBR97}, and so
this means that the measured prestellar lifetime should be a function of
the mean density as well, in qualitative agreement with observational
data \citep{WT_etal07}.

We ignore mergers of collapsing cores in the measurement of
$\tpre$. That is, if two cores appear at times $t_1$ and $t_2 > t_1$
respectively, and they merge at a later time $t_3$ into a new core that
eventually collapses, we take the prestellar lifetime $\tpre$ as
the time elapsed between $t_1$ and the time of density saturation. 
We do this motivated by the fact that the peak densities 
$\npk$ of cores that merge into a new core which collapses later are
seen to rise rapidly even before the merging event, suggesting that
these cores were already on their own independent route to collapse,
rather than the latter being triggered by the merging event. For
instance, the first two cores in R1 (upper panel of Fig.\
\ref{fig:core_peak_dens_evol}, {\it solid} and {\it dashed} lines
respectively) are seen to appear at roughly the same time ($t \sim
1.4$ Myr), and to merge shortly thereafter. Thus, we consider them
together as the single first collapse event in R1.  The third core is seen to
have very similar physical properties to those of the merger, and we
refer to its collapse as the \emph{second} collapse event in R1.  In
Fig.\ \ref{fig:coll_times} we report the prestellar lifetimes measured
for the collapse events in R1 ({\it diamonds}), R2 ({\it triangles})
and R3 ({\it squares}) as a function of their mean densities (which in
turn are $\sim \nthr$).

It is important to remark that the long durations of the collapsed
objects \emph{after} they have collapsed seen in Fig.\
\ref{fig:core_peak_dens_evol} does not imply that the stellar cores
last that long, because no modeling of stellar energy feedback (e.g.,
winds, bipolar outflows, ionizing radiation) is included in the
numerical model, and thus the cores cannot be dispersed after they
produce collapsed objects.

\subsubsection{Number ratios} \label{sec:number_coll}

As mentioned in \S \ref{sec:gral_cons}, we cannot simply count the
number of cores with collapsed objects because it is clear that at the
resolutions we
used their subfragmentation was not appropriately captured. Thus,
instead, we proceed inversely to the standard observational procedure,
and estimate the number ratio of prestellar to stellar cores simply as
the ratio of the prestellar lifetime to the lifetime of the embedded
protostellar phase:
\begin{equation}
\frac{N_\mathrm{pre}}{N_\star}=\frac{\overline{\tau_\mathrm{pre}}}
{\tau_\mathrm{yso}},
\label{eq:Npre/Nstar}
\end{equation}
where $\overline{\tau_\mathrm{pre}}$ is the average of the measured prestellar
lifetimes for the
seven collapse events in the simulations, and $\tyso$ is the duration of the
protostellar stage. We
emphasize that eq.\ (\ref{eq:Npre/Nstar}) allows us to estimate the
number \emph{ratio} of prestellar to stellar cores, but not the actual
numbers, and that this ratio is a function of the threshold density,
since $\tau_\mathrm{pre}$ is.

For us, $\tau_\mathrm{yso}$ is a free parameter because it is outside
the realm of our simulations. As mentioned in \S
\ref{sec:coll_timescales}, no modeling of the feedback from the
protostars such as bipolar outflows, ionization radiation or winds is
included, so there are no energy sources available for dispersing the
cores once they are formed. Thus, we have to consider it as a free
parameter in our study, whose value is to be set on the basis of external
information.

Observationally, $\tyso$ is normally estimated by obtaining the ratio of
the number of embedded objects to that of the T Tauri stars \citep[e.g.,
][]{WLY89, Ken90, Greene_etal94, KH95, Hat06}. Values for
$\tau_\mathrm{yso}$ reported in the literature are $\sim 0.6\pm 0.3$ Myr
\citep[see review by][ and references therein]{WT_etal07}, the spread
being due to both observational constraints and real differences 
among the
observed regions.

We take $\tau_\mathrm{yso}=0.46$ Myr, the mean value
obtained for Perseus by \citet{Hat06}. We base this selection on
two considerations: First, their estimate of $\tau_\mathrm{yso}$
includes corrections for incompleteness of surveys toward very low
masses and multiplicity. Second, we have found that our simulations 
are reasonably
resemblant of a massive ``supercore'' in Perseus, as defined by
\citet{Kirk06} (see \S \ref{sec:simulation} and \S \ref{sec:gral_cons}).

\subsection{Failed cores} \label{sec:failed}

Failed cores in our simulations are less
dense than collapsing cores, and
therefore are not as strongly affected by the limited resolution of our
simulations. Nevertheless, we consider it safer to estimate their number
ratio with respect to the stellar cores through the ratio of masses in
these two kinds of objects. This estimator avoids the uncertainty
associated with the possible failure to capture further subfragmentation
of the cores. However, even this is still not completely straightforward
because we need to take into account temporal effects.

The failed cores ``appear'' at some time and ``disappear'' at a later
time at a given density threshold $\nthr$. Furthermore, their masses are
continuously varying over this time interval, first increasing, reaching
a maximum value, and then decreasing again. In order to obtain an
estimator that is most representative of the number ratio expected to be
found upon a single instantaneous observation, we 
take the ratio of
the temporal averages  of the 
masses of the failed cores  to those of the
stellar cores, weighted by the ratio of their typical durations

\begin{equation}
\frac{N_\mathrm{f}}{N_\star}= \frac{\sum_i \langle
m_i \rangle \tau_i} {\tau_\mathrm{yso}\sum_j \langle m_j \rangle},
\label{eq:failed2stellar}
\end{equation}
where the summations over $i$ and $j$ refer to the failed and stellar
cores respectively, $\langle m_i \rangle$ and $\tau_i$ are respectively
the time-averaged mass and the duration of the $i$-th failed core,
and $\tyso$ is the duration of the protostellar phase discussed in \S
\ref{sec:collapsing}. For the stellar cores, we take $\langle m_j \rangle$
as the average mass over a time interval $\tyso$ after the core has collapsed.

Equation (\ref{eq:failed2stellar}) assumes that 
actual failed cores have,
on average, the same mass as stellar cores. 
However, it is reasonable to expect failed cores to be less  
massive in general than individual stellar cores, 
and in this case the right-hand side of eq.  
(\ref{eq:failed2stellar}) is a lower limit for the failed-to-stellar core
ratio\footnote{See, however, the discussion in \S
\ref{sec:comparison}
concerning the possibility that failed cores 
are hard to detect even at low densities, an effect that would act in the
opposite direction, lowering the observed failed to stellar core ratio.}.
Furthermore, eq. \ref{eq:failed2stellar} also assumes that no 
fragment of a failed core ever ends up collapsing, and no fragment of a
stellar core ends up redispersing. This is a reasonable assumption,
since subfragmentation of the cores is mostly a gravitational 
phenomenon, which occurs after the onset of collapse. Thus, collapsing
cores are expected to fragment into further collapsing units, while
failed cores are not expected to fragment because they do not engage
into collapse. In fact, the failed cores in our simulations generally have
very small, sub-solar  masses (see \S \ref{sec:comparison}).

\subsection{The starless to stellar ratio} \label{sec:svsratio}

Equations (\ref{eq:Npre/Nstar}) and (\ref{eq:failed2stellar}) give us
the prescriptions to estimate the terms on the right-hand side of eq.\
(\ref{eq:starless2stellar}). We can then estimate number ratios with
respect to the total number of cores $N_\mathrm{tot}= N_\mathrm{pre}+
N_\mathrm{f}+N_\star$ in terms of the ratios derived in eqs.\
(\ref{eq:starless2stellar}), (\ref{eq:Npre/Nstar}), and
(\ref{eq:failed2stellar}) as follows:
\begin{eqnarray}
\frac{N_\mathrm{pre}}{N_\mathrm{tot}}&=&\biggl[ \frac{
N_{\star\mathrm{less}} /N_\star} {N_\mathrm{pre}/N_\star} +
\biggl(\frac{N_\mathrm{pre}}{N_\star}\biggr)^{-1}\biggr]^{-1}\\
\frac{N_\mathrm{f}}{N_\mathrm{tot}}&=&\biggl[\frac{N_{\star\mathrm{less}}
/N_\star}{N_\mathrm{f}/N_\star}+\biggl(\frac{N_\mathrm{f}}{N_\star}
\biggr)^{-1}\biggr]^{-1}\\
\frac{N_{\star\mathrm{less}}}{N_\mathrm{tot}}&=&\biggl[1 +
\biggl(\frac{N_{\star\mathrm{less}}}{N_\star}\biggr)^{-1}\biggr]^{-1}\\
\frac{N_\star}{N_\mathrm{tot}}&=&\biggl[1 +
\frac{N_{\star\mathrm{less}}} {N_\star}\biggr]^{-1}
\end{eqnarray}
These fractions are shown in Fig.\ \ref{fig:number_ratios} ({\it left
panel}) as a function of $\nthr$. The {\it right panel} of this figure
shows the same ratios but normalized to the number of \emph{stellar} cores.

\section{Discussion} \label{sec:discussion}

\subsection{Comparison with previous work} \label{sec:comparison}

\begin{figure}
\epsscale{1.0}
\plotone{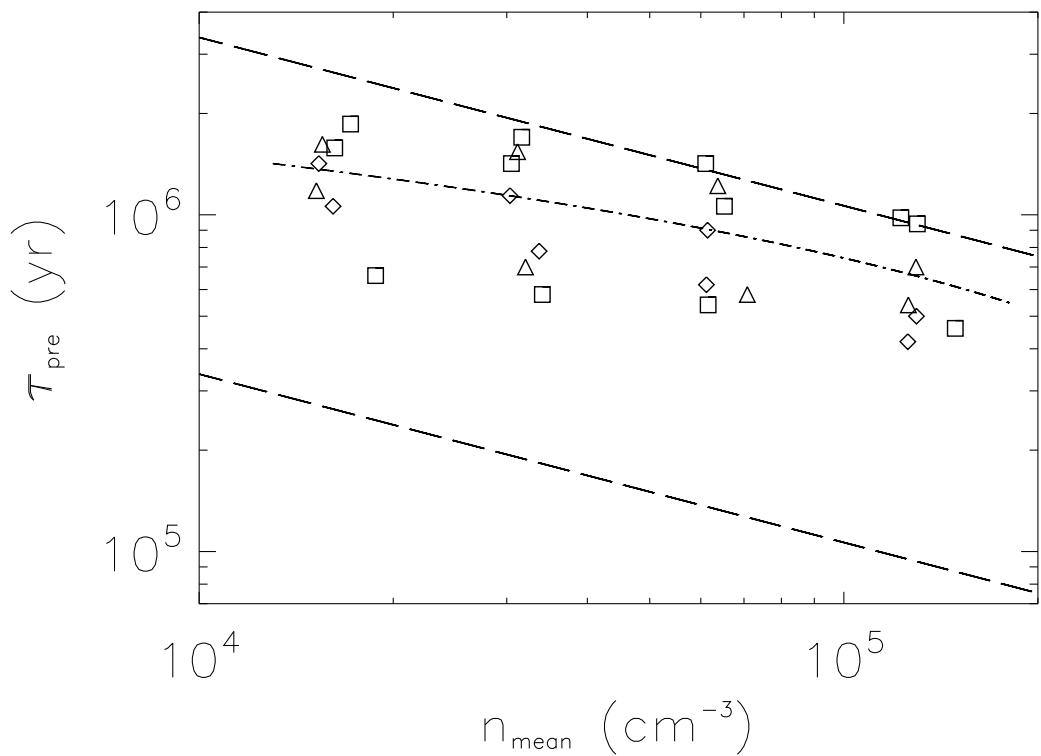}
\caption{Duration of the prestellar stage for the
collapsing cores in the simulations as a function of their
mean density at the time collapse starts being measured. {\it Diamonds},
{\it triangles} and {\it squares} respectively
correspond to the R1, R2 and R3 runs. The loci of
$\tau_\mathrm{pre} = \tff$ and $\tau_\mathrm{pre} = 10~\tff$ is marked by the
lower and upper {\it long-dashed} lines respectively. Also shown is the fit for
the
mean prestellar lifetime as a function of mean density ({\it dash-dotted} line),
which is $\simeq 6~\tff$.  }
\label{fig:coll_times}
\end{figure}

\begin{figure}
\epsscale{1.0}
\plottwo{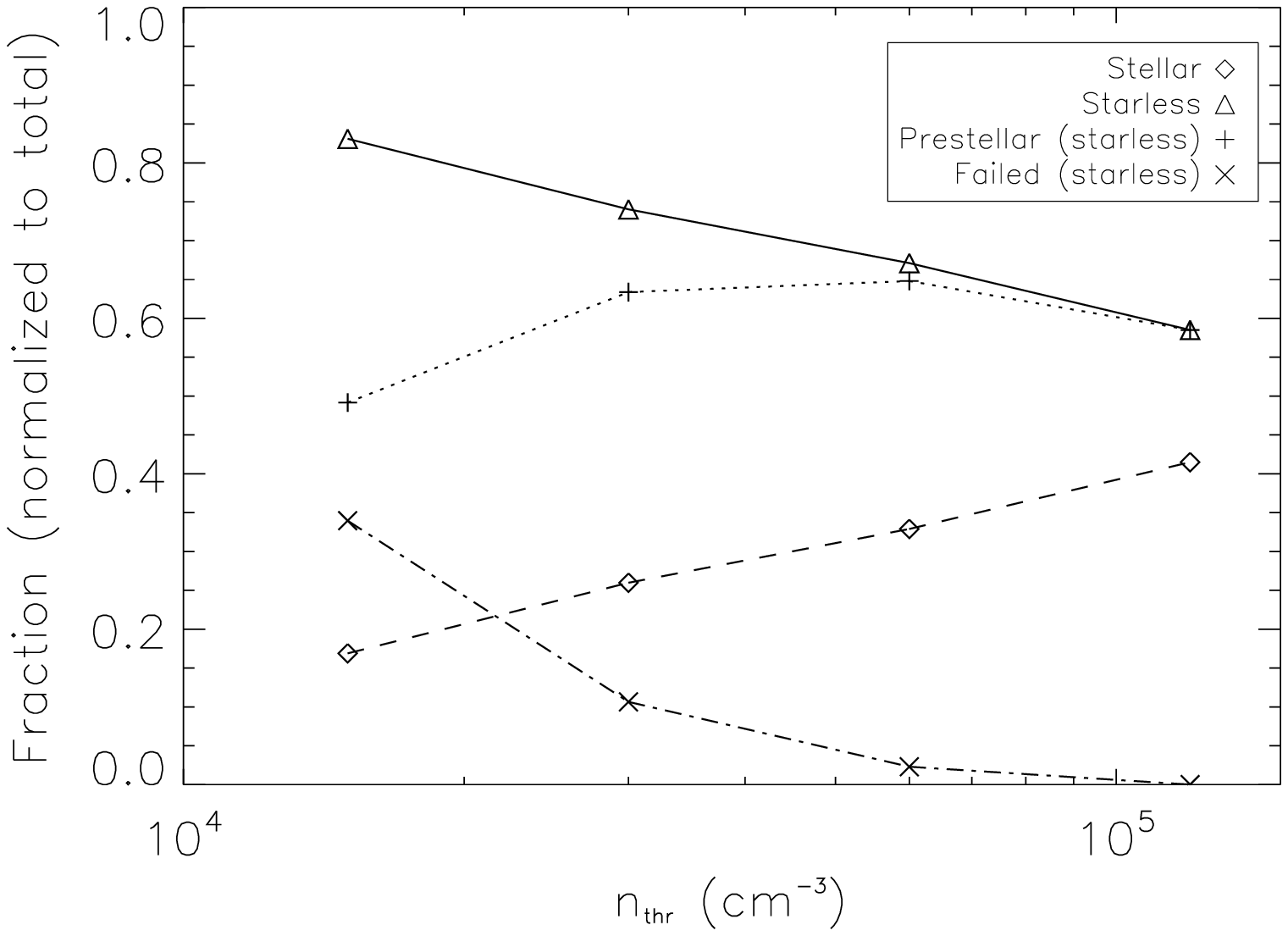}{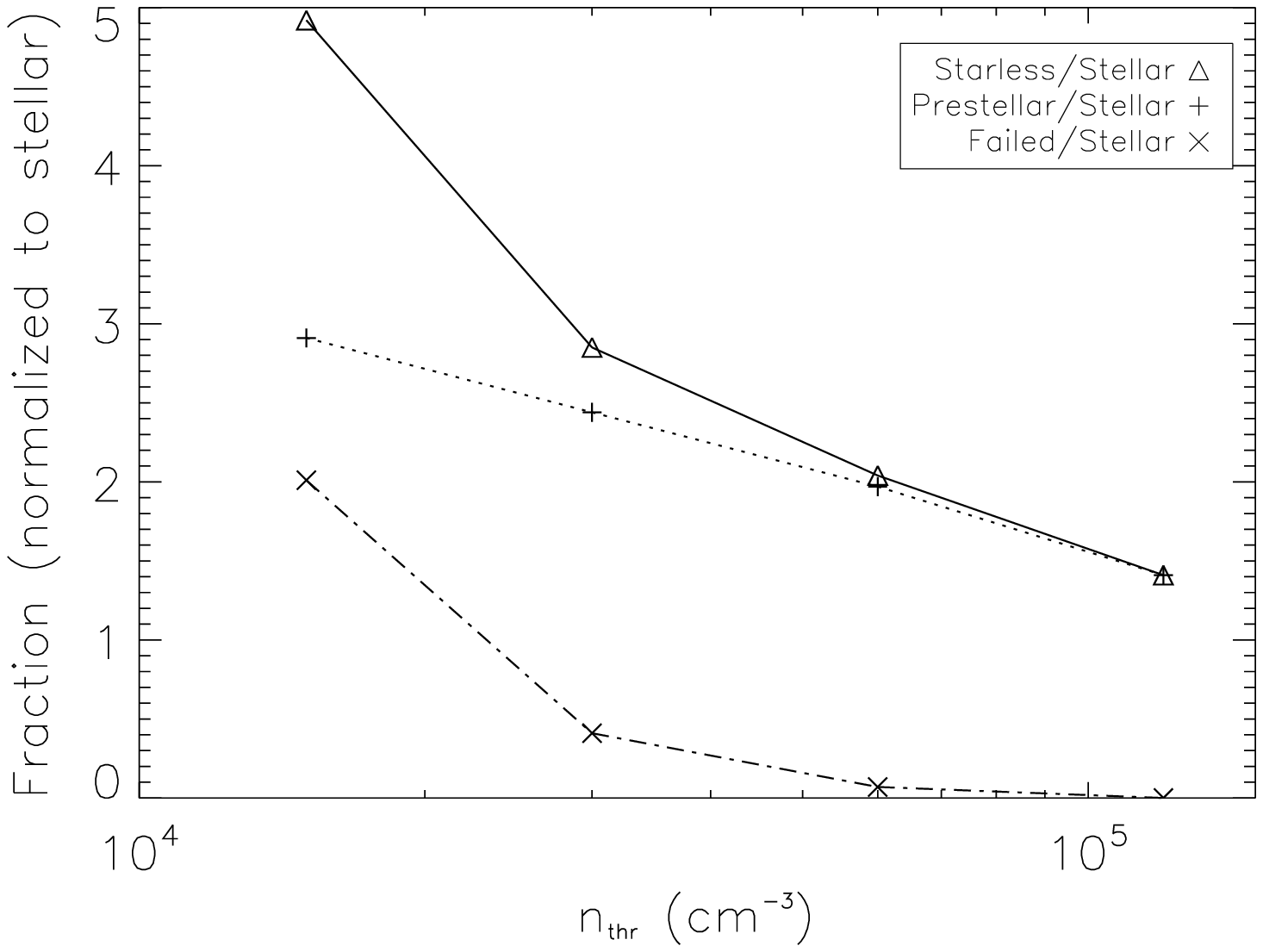}
\caption{{\it Left panel}: Number ratios of the stellar ({\it dashed
line, diamonds}), total starless ({\it solid line, triangles}),
prestellar ({\it dotted line, plus signs}) and failed ({\it dash-dotted
line, crosses}) cores. {\it Right panel}: Number ratios of the total
starless, prestellar and failed cores (same lines and symbols as in the
left panel), normalized to the number of stellar cores.}
\label{fig:number_ratios}
\end{figure}

In addition to the core lifetimes as a function of their mean densities,
Fig.\ \ref{fig:coll_times} shows two lines indicating the locus of
$t=\tff$ and $t=10 \tff$, with $\tff$ defined in eq.\
(\ref{eq:free-fall_time}). This figure can be compared with Figure 2 of
\citet{WT_etal07}. It is seen that, within the uncertainties,
the core lifetimes in our simulations are in good agreement with the
observational values, being in general a few to several times the
free-fall time. However, such a comparison should be taken with care
because in the determination of observational lifetimes, the assumed
values for $\tyso$ span one order of magnitude (from $10^5$ to $10^6$
yr). In any case, one can unambiguously conclude that \emph{the prestellar
core lifetimes in the turbulent scenario are a few times $\tff$}, and not
only $1 \tff$ as is often stated in literature \citep[e.g.,][]{JKirk05,
WT_etal07}.

These results can be understood as a consequence of several factors. 
First, the
cores take some time to be assembled by the external turbulent
compressions before they become gravitationally unstable
\citep{Gomez_etal07}, and even after that, some readjustments of the
mass and some oscillations are expected to occur before gravitational
collapse sets in. Moreover, as pointed out by \citet{Larson69}, the
duration of the actual collapse is longer than the free-fall time
because the thermal pressure gradient is never negligible. So, even though 
the cores in the simulation are formed dynamically, their lifetimes are
fully consistent with observational estimates.

\begin{figure}
\plottwo{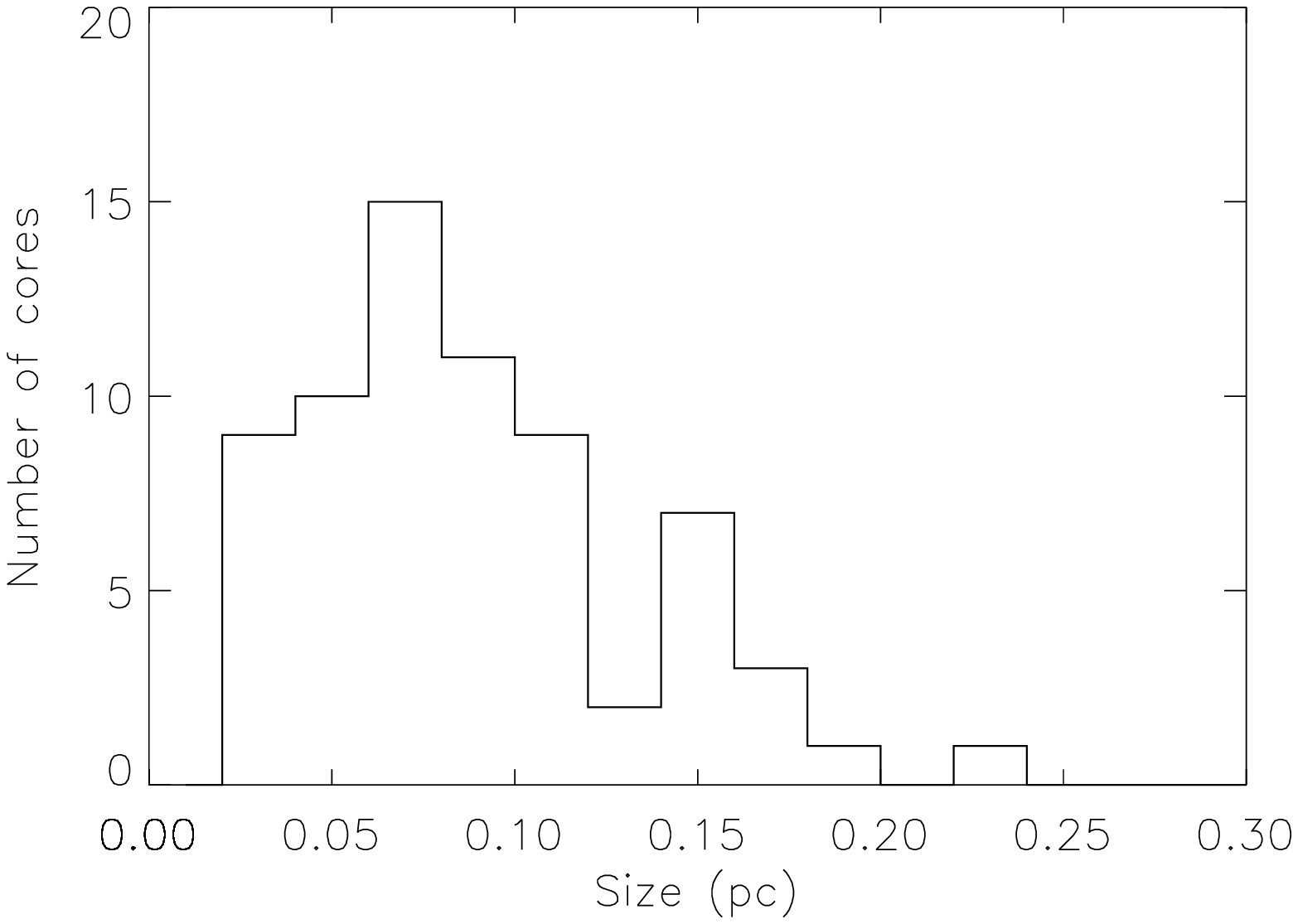}{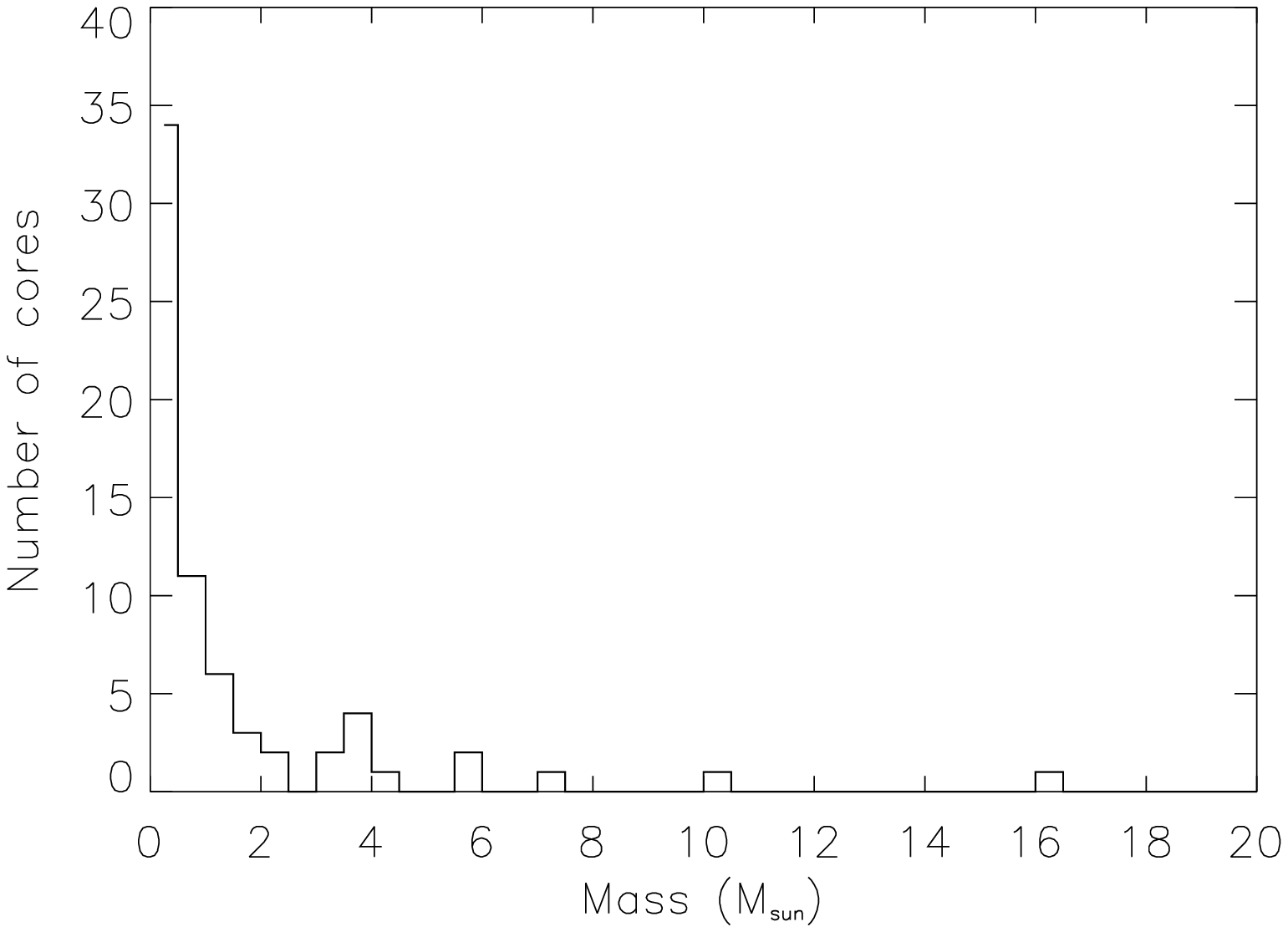}
\caption{{\it Left:} Size histogram of the failed cores at $\nthr=30~n_0 = 1.5
\times 10^4 \pcc$ for R1. {\it Right:} Mass histogram for the same cores.}
\label{fig:size_hist}
\end{figure}

The number ratio of prestellar to stellar cores in our simulations and
the ratio of durations of these stages at low values of $\nthr$ are $\sim
3$ (Fig.\ \ref{fig:number_ratios}, {\it right panel}). The same value is
obtained by \citet{LM99}, who studied an
optically-selected sample of cores, characterized by relatively low mean
densities, $\lesssim 10^4
\pcc$. Note, however, that we obtain a number ratio of
\emph{total} starless (prestellar + failed) to stellar cores of $\sim
5$, larger than the observation of \citet{LM99}. This suggests that
their sample consisted mostly of prestellar cores, possibly because the
failed cores tend to be smaller and therefore less easily
detected. Indeed, Fig.\ \ref{fig:size_hist} ({\it left panel}) shows a
size histogram of the failed cores at $\nthr = 30~n_0 = 1.5 \times 10^4
\pcc$ for R1, and it can be seen that virtually all of them have sizes $\ell <
0.2$ pc, with $\ell \sim 0.07$ pc being a characteristic size; instead, 
\citet{LM99} report a mean size $\ell \simeq 0.2$ pc. Thus, real 
failed cores may be harder to detect observationally, and
therefore not very common in observational surveys, even of low-density
cores. For reference, a histogram of the failed core masses is shown in
the {\it right panel} of Fig.\
\ref{fig:size_hist} as well.

At high thresholds, $\nthr = 1.2\times 10^5 \pcc$, the ratio $N_{\rm
pre}/N_\star \simeq 1.4$. This is consistent with the observed value of $\sim 1$
for this ratio in recent surveys of dense, submm cores
\citep{Hat06,Jor07},
sensitive to $n \sim 10^5 \pcc$. Thus, with a single assumed value of
$\tyso$, our estimated number ratios are in good agreement with
observations at both low and high densities. Moreover, the number of
failed cores at the higher densities goes to zero, suggesting that indeed
submm surveys sample essentially all gravitationally bound
cores, thus tracing the final stages of core evolution prior to
protostar formation.

Our estimated core lifetimes are also in good agreement with the
predictions from the AD-mediated model applied to moderately subcritical
clouds, either in analytical \citep{CB01} or in numerical \citep{NL05}
form. The latter authors report lifetimes 1.5--10 times the local
free-fall time in the collapsing cores, while the lifetimes of our cores
are $\sim 6~\tff$ (see Fig.\ \ref{fig:coll_times}). The similarity between
the mean lifetimes in both cases occurs because, on the one hand,
$\tad$ is only a few times longer than $\tff$ at the densities and
observed values of the mass-to-magnetic flux ratio of molecular cloud
cores while, on the other, the cores in our simulations are not
necessarily in a direct route to collapse when first detected, but
instead undergo a period of 
build up and readjustment, as mentioned above. Therefore,
we conclude that both models give comparable
predictions for the lifetimes of the cores in their observable stages,
with AD not significantly delaying the final stages of contraction and
collapse.

\subsection{Caveats, limitations, and error estimates} \label{sec:caveats}

The study presented in this paper has faced a number of difficulties
comparable to those encountered by observational studies, and in this
sense, its results are only suggestive, rather than conclusive. The
main limitations were:

\medskip
\noindent
1. {\it The limited resolution of the simulations}. This
limitation prevented us from 
correctly following the fragmentation of collapsing cores, and
forced us to estimate the number ratio of prestellar to stellar cores
through the ratio of their durations ($\tau_\mathrm{pre}/\tyso$) rather than
through
direct counting, with $\tyso$ being a free parameter of our study. 
Similarly, the
number ratio of failed to stellar cores had to be estimated in terms of
the ratio of their masses, again due to the inability to resolve
individual stellar cores.

The limited resolution also restricted the number of collapse events to
only $\sim 2$ in each of our runs, although each collapse event clearly
corresponds to the formation of a stellar cluster (i.e., to many collapse events
leading to
individual stars), since the masses of the 
collapsed objects are 50--100 $\Msun$. Unfortunately, we estimate that, in
order to adequately resolve these objects into solar-mass-like cores, an
increase of at least a factor of $50^{1/3}$--$100^{1/3}$, or $\sim$ 4 in
resolution would be required (i.e., resolutions $> 1024^3$), or the
usage of a Lagrangian-type of numerical algorithm (SPH or
adaptive-mesh-refinement), in both magnetic and non-magnetic regimes. We
hope to perform such a study in the near future.

\begin{figure}
\epsscale{0.8}
\plotone{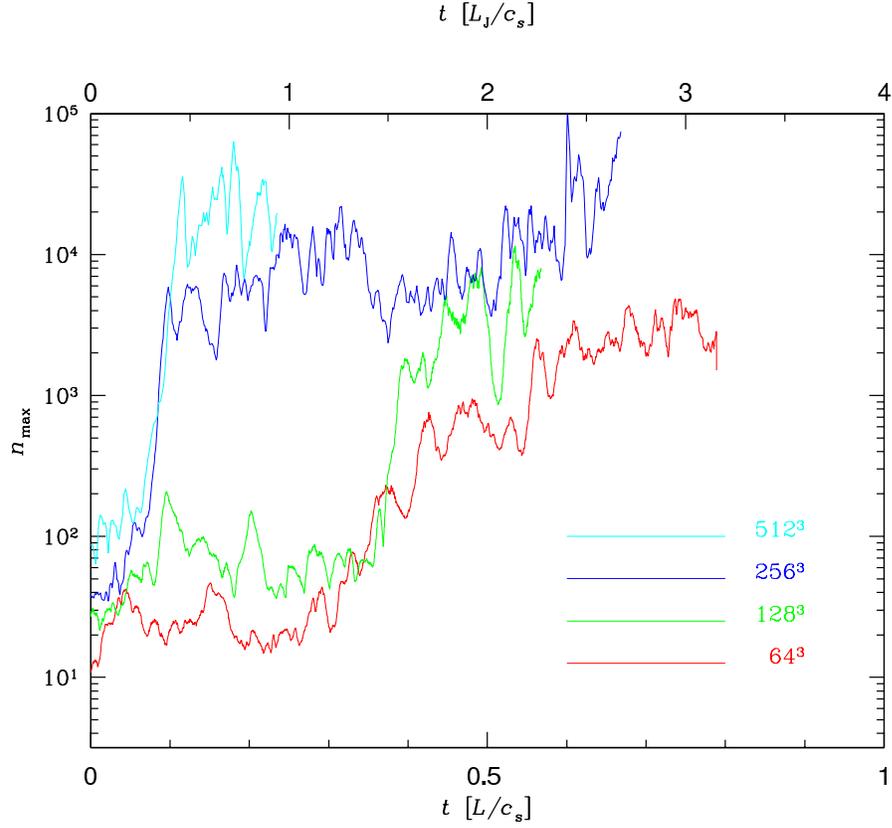}
\caption{Evolution of the maximum density $\nmax$ in four
simulations with the same global parameters (turbulent Mach number $\sim
10$, Jeans number $J=4$) at resolutions $64^3$ ({\it red line}), $128^3$
({\it green line}), $256^3$ ({\it dark-blue line}), and $512^3$ ({\it light-blue
line}). The time axis is given in units of the sound crossing time ({\it
lower axis}), equal to 20 Myr, and of the quantity $\Lj/\cs$ ({\it upper
axis}), which is close to the free-fall time. The density axis is in units of
the mean density in simulations $n_0=500 \pcc$. The first collapse event
in each run is denoted by the
sharp transition from a low mean maximum density level to a high one. The
``collapse time'' of the simulation is defined as the duration of the sharp rise
of $\nmax$ over more than one order of magnitude separating the two
levels. This is well defined in all cases except at $64^3$.}
\label{fig:rhomax_conv_test}
\end{figure}

\begin{figure}
\epsscale{1.0}
\plotone{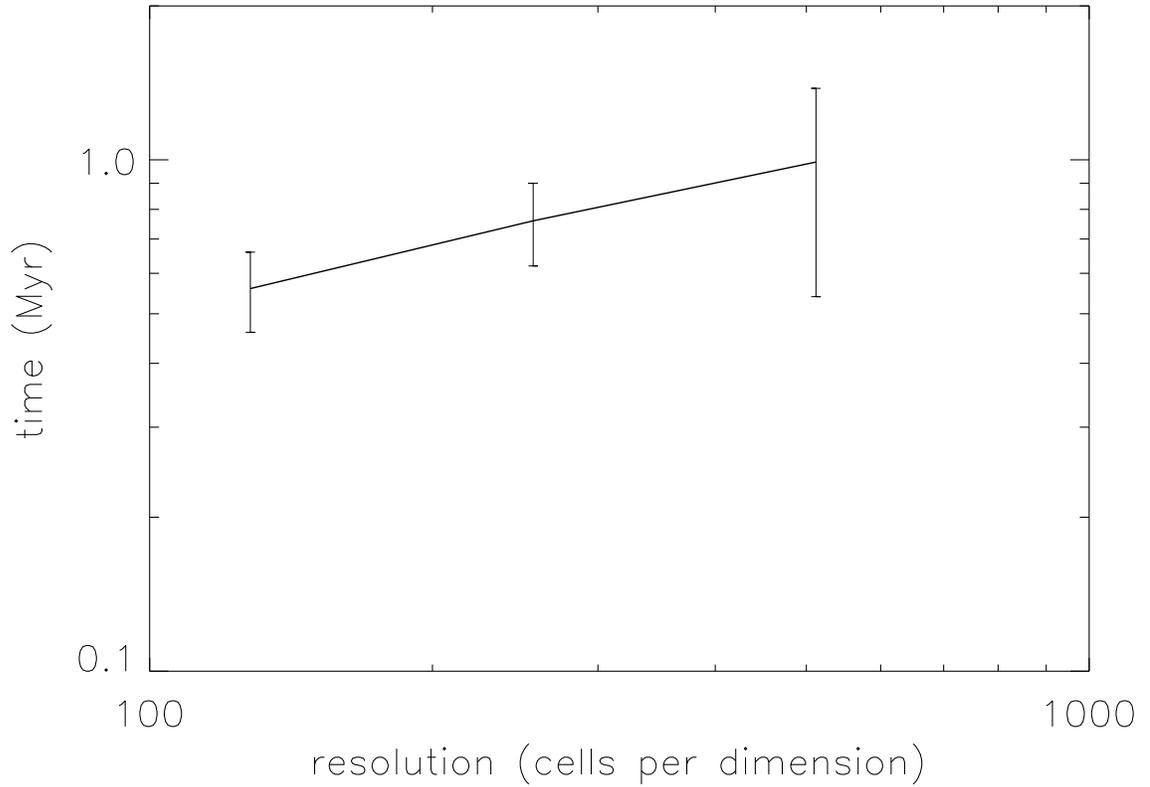}
\caption{Collapse time {\it versus} resolution for the three
highest-resolution runs shown in Fig.\ \ref{fig:rhomax_conv_test}. For
the $128^3$ simulation, this time   
corresponds to the measured duration of the rise of $\nmax$ for
the first collapse, and the error bar denotes the full width of the
rebound around the saturated value. For the $256^3$ and $512^3$
simulations, collapse times correspond to the average of the durations
measured for the collapse events shown in Fig.\
\ref{fig:core_peak_dens_evol}. 
The error bars denote the range of values found.}
\label{fig:colltime_vs_resol}
\end{figure}

The limited resolution might also possibly affect the collapse
times measured in the simulation (see \S
\ref{sec:coll_timescales}), both because of the saturation values of the
density reached by the collapsed objects as well as because of the possible
slowing down of the collapse due to numerical viscosity. To test for this, 
Figures \ref{fig:rhomax_conv_test} and \ref{fig:colltime_vs_resol} 
illustrate the dependence of collapse 
on resolution. Figure \ref{fig:rhomax_conv_test} shows the evolution of the
maximum density $\nmax$ in four simulations with the same global
parameters (turbulent Mach number $\sim 10$, Jeans number $J=4$) at
resolutions $64^3$ ({\it red line}), $128^3$ ({\it green line}), $256^3$
({\it dark-blue line}), and $512^3$ ({\it clear-blue line}). The collapse
in each simulation consists of the sharp transition from a low
typical level of the maximum density ($\nmax \lesssim 100~n_0 = 5 \times
10^4 \pcc$) to a high one ($\nmax \gtrsim 10^3~n_0 = 5 \times 10^5
\pcc$). We define the ``collapse time'' of the simulations
as the duration of the sharp rise of $\nmax$ over more than
one order of magnitude separating the two levels. This is well defined
for resolutions equal or larger than $128^3$. For economy reasons, the
simulation at $512^3$ (i.e., R3) is only run for a short time compared to the
other simulations. In all cases, there is seen to be a rebound from the 
maximum.

Figure \ref{fig:colltime_vs_resol} summarizes the collapse times as a
function of resolution. For the $128^3$ simulation the collapse time 
corresponds directly to the measured duration of the rise of $\nmax$ for
the first collapse, and the error bar denotes the full width of the
rebound around the saturated value. For the $256^3$ (R1) and $512^3$ (R3)
simulations, the collapse times correspond to the average of the durations
measured for the available collapse events. The error bars denote the range of
values found. This figure shows that the collapse time appears to be
independent of resolution for resolutions $128^3$ and above, meaning that
our longer-than-free-fall collapse times are not an artifact of the
resolution.
If anything, Figure \ref{fig:colltime_vs_resol} shows that there is a
slight trend
of the collapse time to {\it increase} with resolution, probably
suggesting that the effect of traversing a larger dynamic range in
density at higher resolution dominates over any slowing down of the
collapse by numerical viscosity at lower resolution.

\medskip
\noindent
2.  {\it The simulations do not include ambipolar diffusion (AD)}. Even
though the simulations were supercritical, in the presence of AD there
would exist the possibility that some failed cores could be ``captured''
by AD if the main agent supporting them is the magnetic energy, since in
this case AD could cause a redistribution of the magnetic field, leaving
the centermost parts of the core with less support than in the case
without AD. This, however, does not appear to be the case. For example,
the longest-lasting failed core in R1, with a lifetime of
$\sim 1.9$ Myr defined at $\nthr = 3 \times 10^4 \pcc$, is clearly
supercritical ($\mu \approx 4.0 \pm 2.1$ times critical, where the
uncertainty comes from the estimation of its ``radius'', since the core
is far from spherical; see Paper I), indicating that its failure to
collapse was due to thermal+turbulent rather than magnetic support.

\begin{figure}
\epsscale{1.0}
\plottwo{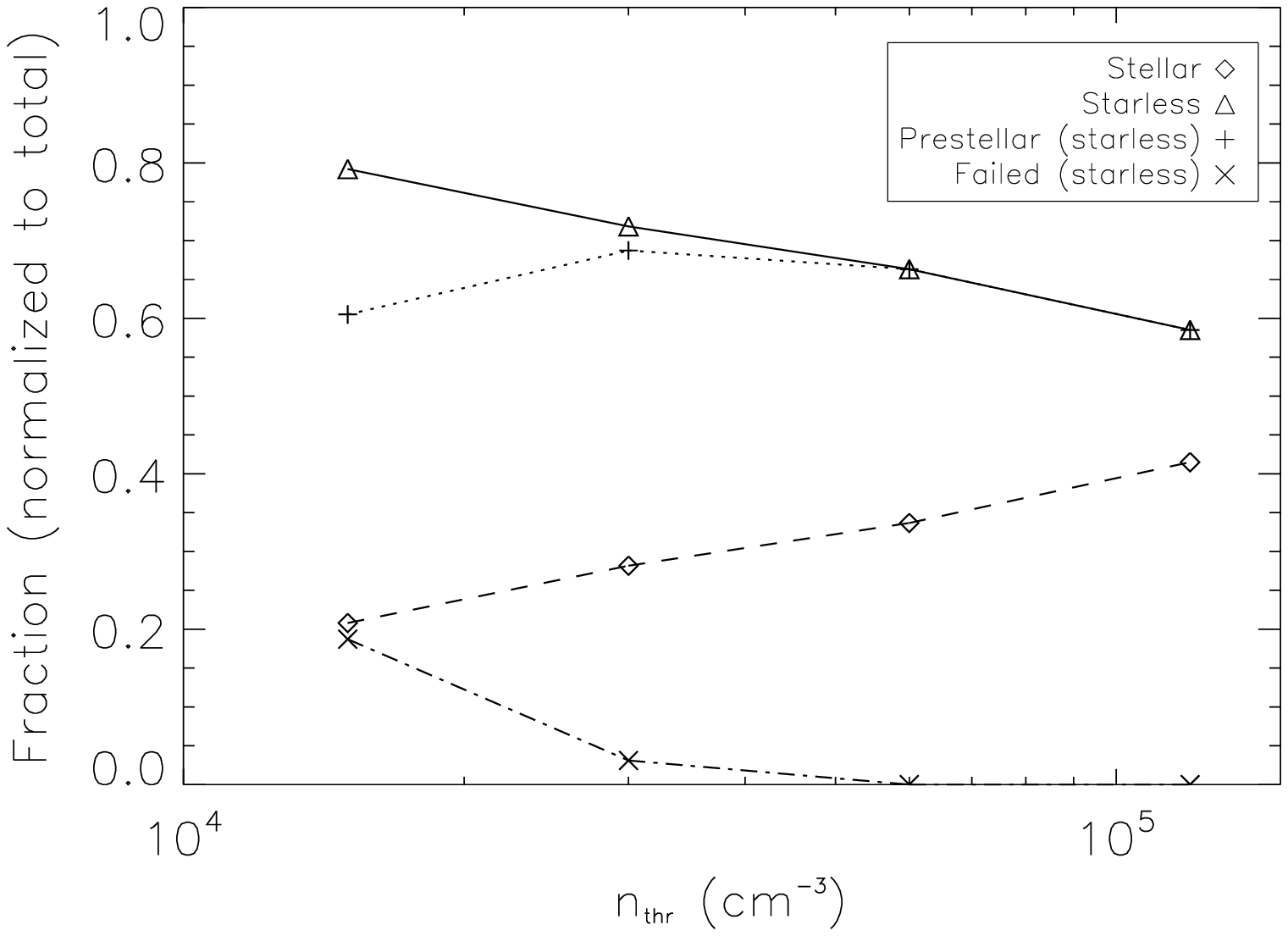}{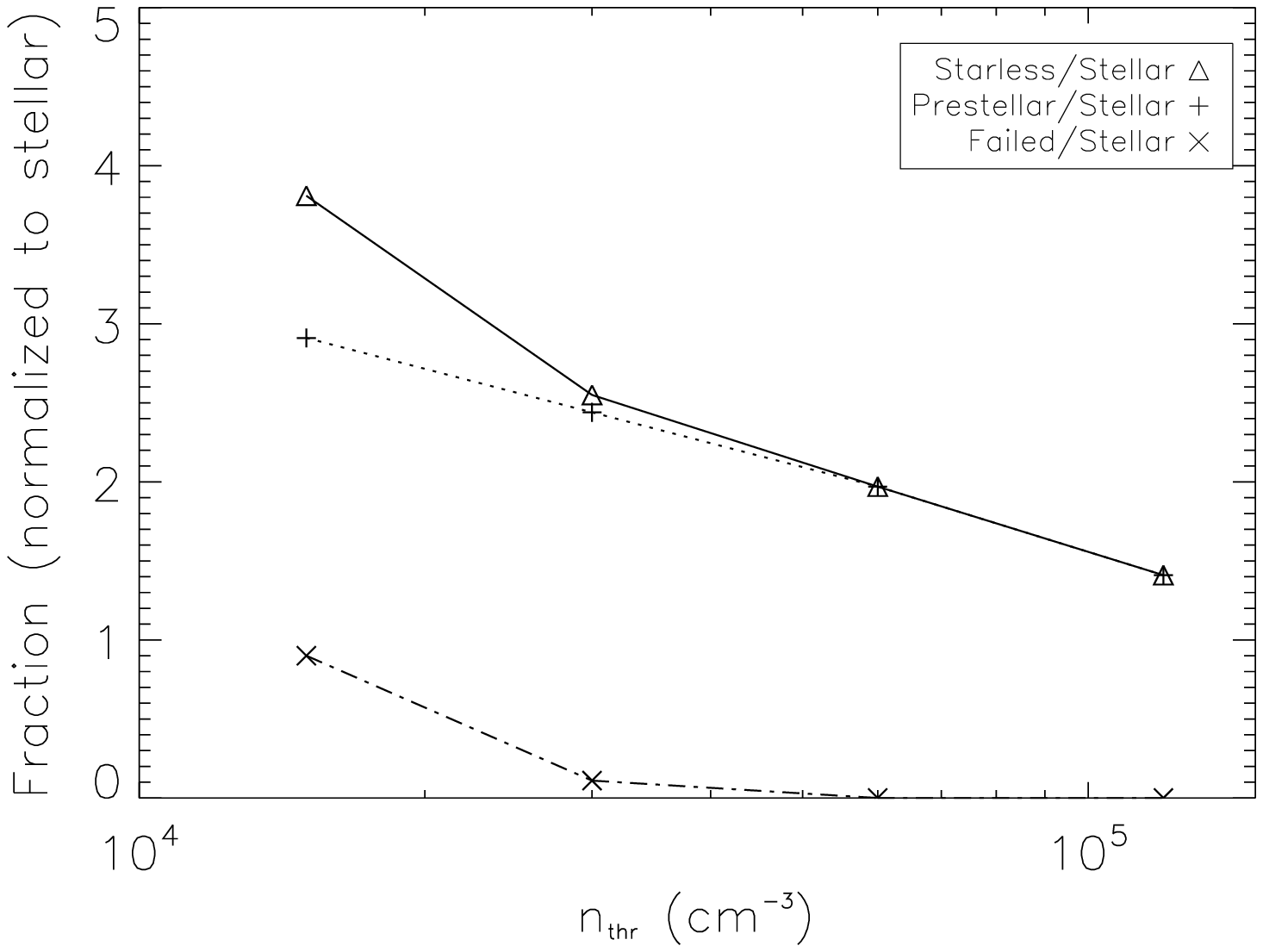}
\caption{Same as Fig.\ \ref{fig:number_ratios} but assuming that all
failed cores with lifetimes $\tau > 0.9$ Myr are ``captured'' by AD and
led to gravitational collapse. The fraction of failed cores is seen to
decrease, but their masses are so small that the fraction of stellar cores
remain essentially unchanged.}
\label{fig:number_ratios_AD}
\end{figure}

In any case, a worst-case estimate of the error committed by the lack of AD
can be made by assuming that all failed cores with lifetimes longer than
a representative AD timescale $\tad$ proceed to collapse rather than
rebound. A reasonable estimate for $\tad$ is $\sim 1.5$ Myr (see
the Appendix in Paper I). Being even more restrictive, we recalculated the
number ratios assuming that the failed cores in R1 with lifetimes $\tau > 0.9$
Myr are ``captured'' by AD (see Fig.\ \ref{fig:number_ratios_AD}). In this
case, we find that the estimated
number ratios of failed cores decrease by a factor of $\sim 2$, but this
does not significantly affect the starless to stellar core ratio,
because the failed cores have very low masses.

\subsection{Conclusions} \label{sec:concl}

In spite of its limitations, the results of the present study are
nevertheless encouraging, as the prestellar lifetimes measured directly from
the density evolution of the cores are in good agreement with
observational determinations. The same is true for the number ratios of
prestellar to stellar cores. This estimation depends on the free
parameter $\tyso$, but a single value for it, chosen as the mean of the
range reported by \citet{Hat06} for Perseus, produces number ratios that
are in good agreement with observational determinations at both high and
low values of the density threshold $\nthr$. The present study thus
suggests that the turbulent model of star formation is not inconsistent
with observational determinations of core lifetimes and the number
ratios of prestellar to stellar cores.

\acknowledgements
We thank Gilberto C. G\'omez for help with IDL visualization. 
We are also grateful for the helpful comments provided by an anonymous referee. 
This work has received partial financial support from CONACYT grant U47366-F
to E. V.-S. The work of J. Kim was supported by the Astrophysical Research
Center for
the Structure and Evolution of the Cosmos (ARCSEC) of Korea Science and
Engineering Foundation (KOSEF) through the Science Research Center (SRC)
program.
The numerical simulations were performed on the Linux cluster
at KAO, with funding from KAO and ARCSEC.

\begin{thebibliography}{}

\bibitem[\BP~et al.\ (1999a)]{BVS99} \BP, J., \VS, E., \& Scalo,
J. 1999a, ApJ, 515, 286

\bibitem[\BP~et al.\ (1999b)]{BHV99} \BP, J., Hartmann, L., \&
\VS, E. 1999b, ApJ, 527, 285

\bibitem[Ballesteros-Paredes et al.(2006)]{BP_etal06}
Ballesteros-Paredes, J., Gazol, A., Kim, J., Klessen, R.~S., Jappsen,
A.-K., \& Tejero, E.\ 2006, \apj, 637, 384

\bibitem[Ballesteros-Paredes \& Hartmann(2007)]{BH07}
Ballesteros-Paredes, J., \& Hartmann, L. \ 2007, Rev. Mex. AA, 43, 123

\bibitem[\BP\ et al.\ (2007)]{BKMV07} \BP, J., Klessen, R., Mac Low,
M.-M., \& \VS, E. 2007, in Protostars and Planets V, ed. B. Reipurth,
D. Jewitt, \& K. Keil (Tucson: Univ. of Arizona Press), 63  

\bibitem[Beichman et al.\ (1986)]{Beich86} Beichman, C. A., Myers,
P. C., Emerson, J. P., Harris, S., Mathieu, R., Benson, P. J., \&
Jennings, R. E. 1986, ApJ, 307, 337

\bibitem[Blitz (1993)]{Blitz93} Blitz, L., 1993, in Protostars and
Planets III, ed. E. H. Levy, \& J. I. Lunine (Tucson: Univ. of
Arizona Press), 125

\bibitem[Ciolek \& Basu (2001)]{CB01} Ciolek, G. E., \& Basu, S. 2001,
ApJ, 547, 272  

\bibitem[Crutcher (1999)]{Crut99} Crutcher, R. M. 1999 ApJ, 520, 706

\bibitem[Elmegreen (1993)]{Elm93} Elmegreen, B. G. 1993, ApJ 419, L29

\bibitem[Elmegreen (2000)]{Elm00} Elmegreen, B. G. 2000, ApJ, 530, 277

\bibitem[Fatuzzo \& Adams (2002)]{FA02}
Fatuzzo, M., \& Adams, F. C. 2002, ApJ, 570, 210

\bibitem[Fiedler \& Mouschovias (1993)]{FM93} Fiedler, R. A., \&
Mouschovias, T. 1993 ApJ, 415, 680

\bibitem[G\'omez et al.\ (2007)]{Gomez_etal07} G\'omez, G. C., \VS, E.,
Shadmehri, M., \& \BP, J. 2007, preprint (arXiv:0705.0559)

\bibitem[Greene et al.\ (1994)]{Greene_etal94} Greene, T. P.,
Wilking, B. A., Andre, P., Young, E. T., \& Lada, C. J. 1994, ApJ, 434, 614

\bibitem[Hartmann et al.\ (2001)]{HBB01} Hartmann, L.,
Ballesteros-Paredes, J., \& Bergin, E. A. 2001, ApJ, 562, 852

\bibitem[Hatchell et al.\ (2007)]{Hat06} Hatchell, J., Fuller, G. A.,
Richer, J. S., Harries, T. J., \& Ladd, E. F. 2007, A\&A, 468, 1009

\bibitem[Heitsch et al.\ (2004)]{Heitsch_etal04} Heitsch, F., Zweibel,
E. G., Slyz, A. D., \& Devriendt, J. E. G. 2004, ApJ, 603, 165

\bibitem[Heyer \& Brunt (2004)]{Heyer_Brunt04} Heyer, M.~H., \& Brunt,
C.~M.\ 2004, \apjl, 615, L45

\bibitem[Heyer \& Brunt (2007)]{Heyer_Brunt07} Heyer, M.~H., \& Brunt,
C.~M.\ 2007, in IAU Symposium 237, Triggered Star Formation in a
Turbulent ISM, ed. B. Elmegreen \& J. Palous (Cambridge: Cambridge
University Press), 9

\bibitem[Jijina et al.\ (1999)]{JMA99} Jijina, J., Myers, P. C., \&
Adams, F. C. 1999, ApJS, 125, 161

\bibitem[J$\o$rgensen et al.\ (2007)]{Jor07} J\o rgensen, J. K.,
Johnstone, D., Kirk, H., \& Myers, P. C. 2007, ApJ, 656, 293

\bibitem[Kenyon et al.\ (1990)]{Ken90}
Kenyon, S. J., Hartmann, L. W., Strom, K. M., \& Strom, S. E.
1990, AJ, 99, 869

\bibitem[Kenyon \& Hartmann (1995)]{KH95} Kenyon, S. J., \& Hartmann,
L. 1995, ApJS, 101, 117

\bibitem[Kim et al.\ (1999)]{Kim_etal99} Kim, J., Ryu, D., Jones, T. W.,
\& Hong, S. S. 1999, ApJ, 514, 506

\bibitem[Kirk et al.\ (2006)]{Kirk06} Kirk, H., Johnstone, D., \& Di Francesco, J. 2006, ApJ, 646, 1009

\bibitem[Kirk et al.\ (2005)]{JKirk05} Kirk, J. M., Ward-Thompson, D.,
\& Andr\'e, P. 2005, MNRAS, 360, 1506

\bibitem[Larson(1969)]{Larson69} Larson, R.~B.\ 1969, \mnras,
145, 271

\bibitem[Larson(1981)]{Larson81} Larson, R.~B.\ 1981, \mnras,
194, 809

\bibitem[Lee \& Myers (1999)]{LM99} Lee, C. W., \& Myers, P. C. 1999,
ApJS, 123, 233

\bibitem[Mac Low \& Klessen (2004)]{MK04} Mac Low, M.-M., \& Klessen,
R. S. 2004, Rev. Mod. Phys., 76, 125

\bibitem[Maddalena \& Thaddeus (1985)]{Maddalena_Thaddeus85}
Maddalena, R. J., \& Thaddeus, P. 1985, ApJ, 294, 231

\bibitem[Mouschovias (1976)]{Mou76} Mouschovias, T. C. 1976, ApJ, 207, 141

\bibitem[Mouschovias (1991)]{Mou91} Mouschovias, T. Ch. 1991, in The
Physics of Star Formation and Early Stellar Evolution, ed. C. J. Lada,
\& N. D. Kylafis (Dordrecht: Kluwer), 449

\bibitem[Nakamura \& Li (2005)]{NL05} Nakamura, F., \& Li, Z.-Y. 2005,
ApJ, 631, 411

\bibitem[Onishi et al.\ (2002)]{Oni02} Onishi, T., Mizuno, A., Kawamura,
A., Tachihara, K., \& Fukui, Y. 2002, ApJ, 575, 950

\bibitem[Padoan \& Nordlund (2002)]{PN02}
Padoan, P., \& Nordlund, \AA\ 2002, ApJ, 576, 870

\bibitem[Shu et al.\ (1987)]{SAL87} Shu, F. H., Adams, F. C., \& Lizano,
S. 1987, ARA\&A, 25, 23

\bibitem[Stone et al.\ (1998)]{Stone98} 
Stone, J. M., Ostriker, E. C., \& Gammie, C. F. 1998, ApJ, 508, L99

\bibitem[Taylor et al.\ (1996)]{TMW96} Taylor, S. D., Morata, O., \&
Williams, D. A. 1996, A\&A, 313, 269

\bibitem[Truelove et al.\ (1997)]{True_etal97} Truelove, J. K., Klein,
R. I., McKee, C. F., Hilliman, J. H. II., Howell, L. H., \& Greenough,
J. A. 1997, ApJ, 489, L179

\bibitem[\VS\ et al.\ (1997)]{VBR97} \VS, E., \BP, J., \& Rodr\'iguez,
L. F. 1997, ApJ, 474, 292

\bibitem[\VS\ et al.\ (2000)]{VS_etal00} V\'azquez-Semadeni, E.,
Ostriker, E. C., Passot, T., Gammie, C., 
\& Stone, J. 2000, in Protostars and Planets IV, ed. V. Mannings,
A. Boss, \& S. Russell (Tucson: Univ.\ of Arizona Press), 3

\bibitem[\VS\ et al.\ (2005a)]{Paper I} \VS, E., Kim, J., Shadmehri, M., \& 
\BP, J. 2005a, ApJ, 618, 344 (Paper I)

\bibitem[\VS\ et al.\ (2005b)]{VKB05} \VS, E., Kim, J., \&
\BP, J. 2005, ApJ, 630, L49

\bibitem[\VS\ (2007)]{VS07} \VS\ 2007, in IAU Symp. 237, Triggered Star
Formation in a Turbulent ISM, ed. B.G. Elmegreen, \& J. Palous
(Dordrecht: Kluwer), 50

\bibitem[Ward-Thompson et al.\ (2007)]{WT_etal07} Ward-Thompson, D.,
Andre, P., Crutcher, R., Johnstone, D., Onishi, T., \& Wilson, C. 2007, in
Protostars and Planets V, ed. B. Reipurth, D. Jewitt, \& K. Keil
(Tucson: Univ.\ of Arizona Press), 33

\bibitem[Wilking et al.\ (1989)]{WLY89} Wilking, B. A., Lada, C. J., \&
Young, E. T. 1989, ApJ, 340, 823

\bibitem[Williams et al.(1994)]{Williams_etal94} Williams, J.~P., de
Geus, E.~J., \& Blitz, L.\ 1994, \apj, 428, 693


\end{thebibliography}
\end{document}